\pdfoutput = 1
\RequirePackage{nag}
\RequirePackage{amssymb}
\RequirePackage{amssymb}
\documentclass[runningheads]{raa}
\usepackage[pagebackref=true,bookmarks=false, colorlinks = true, linkcolor = green, anchorcolor = red, citecolor = blue, filecolor = red, urlcolor = red]{hyperref}
\usepackage{graphicx,times}
\usepackage{natbib}
\usepackage{multirow}

\bibpunct{(}{)}{;}{a}{}{,}

\begin{document}

   \title{First Time-Resolved CCD Photometry and Time-series Analysis of NSV
   13601}
   \volnopage{Vol.0 (20xx) No.0, 000--000}\par\vspace{\baselineskip}
   \setcounter{page}{1}
   \author{Virabhadrasinh A. Gohil and Sachchidanand Prakash Bhatnagar}
   \institute{Department of Physics, M. K. Bhavnagar University, Bhavnagar -
   364001, Gujarat, India; {\it spb@mkbhavuni.edu.in}\newline\newline
   \vs\no{\small Received~~2019~~June 24; accepted~~2019~~October 15}}
   \date{}
   \abstract{We present the first results of a time-resolved CCD photometric and
   time-series analysis of NSV 13601, a variable star of Pegasus constellation.
   The 14" SCT of Maharaja Krishnakumarsinhji Bhavnagar University(MKBU) was used
   for observations. Analysis was performed both in Vstar and Period04 software
   and compared their results. The main results are as follows: From data
   analysis, it's period, $\sim$77.784 d(MKBU Data V)(VStar), $\sim$77.632 d(MKBU
   Data Ic)(VStar), $\sim$77.058 d(MKBU Data V)(Period04) and  $\sim$49.560 d(MKBU
   Data Ic)(Period04). It's V and I mean magnitudes for MKBU Data are 12.203 (V)
   and  11.292 (I) mag respectively. We confirm it as a variable star.
   }

   \maketitle

\section{Introduction}
\label{sect:intro}

NSV
13601\footnote{\texttt{https://www.aavso.org/vsx/index.php?view=detail.top\&oid=52225}}(RA:
21:13:01.24, DEC: +18:56:29.2), an eclipsing binary type variable star in the
Pegasus constellation.\par\vspace{\baselineskip}

\noindent Many astronomical databases were searched for collecting details about
this star, the details are given as below,\par\vspace{\baselineskip}

\noindent On VSX
database\footnote{\label{vsx}\texttt{https://www.aavso.org/vsx/}}, It is
designated as Suspected Variable with Mag. (11.31 - ? V) and having variability
type as E. It's other names are 2MASS J21130123+1856292, GSC 01658-00370 and TYC
1658-370-1.\par\vspace{\baselineskip}

\noindent On GCVS
database\footnote{\texttt{http://www.sai.msu.su/gcvs/cgi-bin/search.htm}}, it's
type is E:, magnitude range is Max.: ~13. and Min.: \textless
~13.5.\par\vspace{\baselineskip}

\noindent Many other astronomical databases like, CDS
portal\footnote{\texttt{http://cdsportal.u-strasbg.fr/}}, CRTS
DR3\footnote{\texttt{https://crts.iucaa.in/CRTS/}}, The Catalina
survey\footnote{\texttt{http://crts.caltech.edu/}},
Hipparcos\footnote{\texttt{https://www.cosmos.esa.int/web/hipparcos}}, The
Hipparcos Main
Catalog\footnote{\texttt{https://heasarc.gsfc.nasa.gov/W3Browse/all/hipparcos.html}},
DASCH(apass) Catalog\footnote{\texttt{http://dasch.rc.fas.harvard.edu/index.php}},
gaia\footnote{\texttt{https://www.cosmos.esa.int/web/gaia/dr2}},
Hipparcos-2\footnote{\texttt{https://www.cosmos.esa.int/web/hipparcos/hipparcos-2}},
ASAS-SN\footnote{\texttt{https://asas-sn.osu.edu/variables}} and
AAVSO\footnote{\texttt{https://www.aavso.org/data-download}} were searched, but no
relevant information was found about NSV 13601.\par\vspace{\baselineskip}

\noindent After collecting all these information about Target Star, finder chart
was prepared from AAVSO star chart plotter along with it's photometry table,
comparison and check star were selected from the table nearest to the variable
star and with magnitude very close to target variable star. All the three stars
were in the same CCD frame.\par\vspace{\baselineskip}

\noindent The standard CCD differential photometry was performed and details about
variable star, comparison star and check star is given in
Table-2. Also, MKBU CCD (V)image is given in  Fig. 1.\par\vspace{\baselineskip}


\noindent Total of 34 nights worth of data was collected and total number of
frames were 170 each in (V) and  (Ic) filters
respectively.\par\vspace{\baselineskip}

\noindent Light curves were prepared for MKBU Data-V and MKBU Data-Ic as shown in
Fig. 2 and 5 respectively,  for further time-series
analysis.\par\vspace{\baselineskip}

\section{Methodology}
\label{sect:meth}

\subsection{Instrumentation}

All the photometric data analyzed for this paper were obtained using 14" optical
telescope; Celestron Schmidt-Cassegrain reflector telescope mounted at MK
Bhavnagar University Observatory, Bhavnagar (Lat: 21.7542° N ; Long: 72.1304° E),
India.\footnote{\texttt{https://www.google.com/maps/place/Bhavnagar+Observatory/@21.7543468,72.1303431,19z\\/data=!4m5!3m4!1s0x0:0x30504b9860ad6057!8m2!3d21.7542085!4d72.1303511?hl=en}}(\citealt{Bhatnagar+etal+2001})\par\vspace{\baselineskip}

 \noindent The  automated Observatory contains a 14" Celestron Schmidt-Cassegrain
 reflector type Telescope (D = 355.6mm, F = 3910mm, f/11) mounted on a  Equatorial
 mount. Telescope has an SBIG ST-7XME CCD Camera(Table-1) along with SBIG CFW-8A
 filter wheel with Johnson-Cousins \textit{UBVRI} photometric
 filters\footnote{\texttt{https://en.wikipedia.org/wiki/Photometric\_system}}(\citealt{Bessell+1990}).
 The telescope has two stepper motors (Aerotech Model 310SML3) for RA and DEC axes
 which are controlled by micro-stepping stepper motor translators (Aerotech
 DM8010). The whole system was indigenously developed in collaboration with IUCAA,
 Pune\footnote{\texttt{https://www.iucaa.in/}} (Inter University Center for
 Astronomy and Astrophysics). For telescope control SCOPE(by Mel Bartel) was used
 in conjunction with Cartes Du Ciel. There is an USB connection between CCD and PC
 for data as well as commands. Astronomy data is locally stored in  PC. A GPS
 receiver is also connected with PC which periodically updates clocks of the
 computers.\par\vspace{\baselineskip}

\noindent The stepper motors are mechanically coupled to the telescope through a
friction drive of ratio 1:24. Thus the movement of telescope is finely controlled
by the stepper motor to the tune of 25000 steps per rotation in both the axes
resulting in about 2.16 arc sec per step accuracy. Friction drive was selected to
avoid backlash. The friction drive wheels of 1 inch and 24 inch diameter are
machined to an accuracy of a few micron and are nitride
hardened.\par\vspace{\baselineskip}

\noindent Only V and I, Standard Johnson-Cousins filters were used for this study
as other filters were found to have deteriorated by environmental effects so their
data were not used.\par\vspace{\baselineskip}

\noindent Technical specifications of MKBU telescope with ST-7XME CCD camera and
focal reducer used  are given in
table-1.\footnote{\texttt{https://www.qdigital-astro.com/calculator}}\par\vspace{\baselineskip}

\begin{table}[ht]
\resizebox{\textwidth}{!}{
\begin{minipage}{15cm}
\centering
\begin{tabular}{|l|l|}

 \hline
 CCD camera:&SBIG ST-7XME\\
 \hline
 Telescope:&Celestron 14" f/11\\
 \hline
 Reducer:&0.63xReducer\\
 \hline
 Pixel Binning:&1x1\\
 \hline
 Image scale:&0.75 arsec/pixel\\
 \hline
 Focal length:&2463mm\\
 \hline
 Focal ratio:&76.93\\
 \hline
 field of view of CCD camera:&9.6' x 6.4'(arcmin, width x height)\\
 \hline
\end{tabular}
\end{minipage}}
\caption{\textbf{Technical details of MKBU Telescope system with focal reducer}}
\end{table}

\subsection{Observations and data reduction}

MaxIm DL photometric image analysis software was used for our data reduction work.
All time-series astronomical images were first calibrated for basic noise removal
and then all continuous sequences were aligned and  photometry performed on
them.\par\vspace{\baselineskip}

\noindent For flat fielding twilight images of sky were captured and used daily.
After all the images were captured, master files were made for each of these and
used  for final calibration part.\par\vspace{\baselineskip}

\noindent In MaxIm DL, all the flats, darks and bias files were combined into
single standard master file according to different photometry filters. For
calibration, these master files were used. The process involved using the object
star’s light image and  perform calibration using the master calibration files in
MaxIm DL. This way all the light data images were calibrated for each observation.
Subsequently these calibrated light images were aligned using the alignment tool
of the MaxIm DL. These aligned images were integrated to get better signal-noise
ratio.\par\vspace{\baselineskip}

\noindent These calibrated and integrated light images were used in differential
photometry tool of MaxIm DL, by selection of each individual object star,
reference star and check stars. By providing proper inputs to the MaxIm DL this
results in a Magnitude Vs JD(Jullian Day) graph in CSV or AAVSO Text format file.
The time accuracy is as good as provided by GPS time
syncing.\par\vspace{\baselineskip}

 \noindent An object star, one check stars(standard star) and one comparison star
 were selected to perform differential photometry(\citealt{Rodrigez-Gil+2005}).
 Standard observatory procedures were used with minimum air-mass
 considerations.\par\vspace{\baselineskip}

\noindent For data reduction(calibration), flats field frames, dark frames and
bias frames as well as light
frames were  set into MaxIm DL configuration which then collects each of these
automatically according to set exposure time and other Johnson-Cousins
\textit{UBVRI} filter value settings.
All the light curve analysis were performed in Vstar, to double check the results,
Period04 (a standard astronomy time-series analysis package) was  also
used.\par\vspace{\baselineskip}

\noindent Aladin\footnote{\texttt{https://aladin.u-strasbg.fr/}} was used to
compare the captured CCD image with the standard field
image.\par\vspace{\baselineskip}

\subsubsection{Differential Photometry}

For standardization of CCD photometric method and to calibrate the whole system, 5
known variable stars were studied. One of which was presented in the
paper(\citealt{Gohil+Bhatnagar+2019}). This section presents the basic CCD
differential photometry performed on suspected variable NSV
13601.\par\vspace{\baselineskip}

\noindent For finding suspected variable star
GCVS\footnote{\texttt{http://www.sai.msu.su/gcvs/gcvs/}} and VSX$^{\ref{vsx}}$
databases were used.  NSV 13601, a  suspected variable star was shortlisted based
on it's closeness to zenith at the observation location and time.  NSV 13601 was
selected  due to it's magnitude less than 15, suitable for observing from small
telescope located in semi urban area.\par\vspace{\baselineskip}

\noindent All the flats (flat CCD image frames) were also taken in different
filters daily at the dusk time when the sky was evenly lit. The CCD's internal
Peltier cooler was kept at it's minimum temperature (around -20 $^{\circ}$C)
during all the observations to reduce the thermal noise.\par\vspace{\baselineskip}

\noindent The target observation photometric sequences used were optimum for
differential photometry(\citealt{Miles+1998}). The sequence of observation
were,\par\vspace{\baselineskip}

\noindent Bias -\textgreater Dark -\textgreater \textquotedblleft Object Star +
Comparison Star + Check Star\textquotedblright-\textgreater Dark-\textgreater Bias
(For each Filter)\par\vspace{\baselineskip}

\noindent The \textquotedblleft Object Star + Comparison Star + Check
Star\textquotedblright~were chosen so that, they were near to each other and could
be located in the same CCD image frame.\par\vspace{\baselineskip}

\noindent For aperture photometry, the aperture, annulus and gap were chosen so
that the aperture only contained the star's light and the annulus contained the
background.\par\vspace{\baselineskip}

\noindent From the image thus obtained, the intensity was differentially derived
by Photometry tool of MaxIm
DL\footnote{\texttt{\label{diffltd}https://diffractionlimited.com/}} for the
target using the known comparison and standard stars. The light curve were
obtained for the target star from each combined CCD image generated at different
times.\par\vspace{\baselineskip}

\noindent Data from V and Ic photometric filter was used as the data from other
filters was found to be unreliable due to their environmental deterioration. The
standard extinction correction could not be carried out in the conventional
photometric way. Therefore for atmospheric corrections, it became  advantageous to
observe only when the object star, comparison star and check star were at the
zenith. The CCD differential photometry takes care of the corrections.  Moon less
clear (photometric) nights were selected for
observations.\par\vspace{\baselineskip}

\noindent The data labeled as  MKBU is taken at MKBUs
observatory\footnote{\texttt{\label{bhavuni}https://mkbhavuni.edu.in/mkbhavuniweb/}}.\par\vspace{\baselineskip}

\noindent \textbf{Suspected Variable Stars}\par\vspace{\baselineskip}

\noindent The data presented here was captured at the available beginning of this
study($\sim$2014). Since then some additional information on variability have
become available. Details given for NSV 13601.\par\vspace{\baselineskip}

\begin{table}[ht]
\resizebox{\textwidth}{!}{
\begin{minipage}{21 cm}
\centering
\begin{tabular}{|c|c|c|c|c|c|c|c|c|}

 \hline
 \textbf{Star type}& \textbf{AUID}&
 \multicolumn{2}{|c|}{\textbf{J2000.0}}&
 \multicolumn{5}{c|}{\textbf{Magnitudes}}\\

 &&($RA$)&($DEC$)&($B$)&($V$)&($B-V$)&($Rc$)&($Ic$)\\

 \hline
 Object Star&000-BCS-770&21:13:01.24&18:56:29.20&-&11.31-?&-&-&-\\
 \hline
 Check Star&000-BCS-776&21:13:11.29&18:57:12.1&11.665(0.155)\^*&11.386(0.026)\^@&0.279(0.157)&-&10.833(0.143)\^\$\\
 \hline
 Comparison Star-1&000-BCS-777&21:13:13.29&18:58:03.7&-&11.875(0.137)\^\$&-&-&11.144(0.217)\^\$\\
 \hline
 Comparison Star-2&000-BCS-764&21:12:37.32&19:02:54.9&12.478(0.257)\^*&11.044(0.023)\^@&1.434(0.258)&-&9.667(0.126)\^\$\\
 \hline

\end{tabular}
\begin{flushleft}
Here,  * = Tycho-2 , \$ = TASS , @ = ASAS3
\end{flushleft}
\end{minipage}}
\caption{\textbf{Object, check and comparison star table:(NSV 13601)}}
\end{table}

\begin{figure}

\centering{\includegraphics[width=10cm]{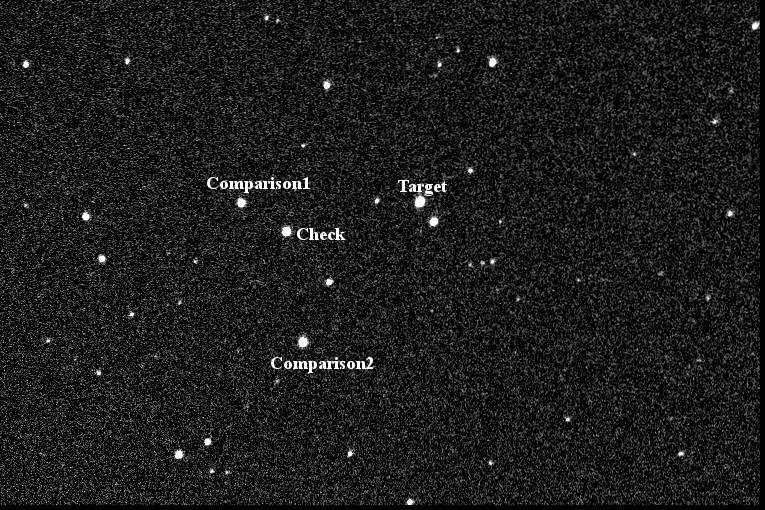}}

\caption{Object, check and comparison star field-CLEAR Filter: 16-04-2014\label{Fig1}}

\end{figure}

\newpage

%
%
%
%
%

\subsubsection{Time-series analysis}

Time-series analysis on the data collected at MKBU observatory$^{\ref{bhavuni}}$
and the results derived, are presented in this section. This time-series analysis
was mainly done in Vstar(AAVSO)
software.\footnote{\texttt{http://www.citizensky.org/content/vstar}} Light curves
was generated by MaxIm DL$^{\ref{diffltd}}$ and directly used as input into VStar
software. Time-series analysis was performed to find periods of NSV 13601.
Initially Gemini constellation was studied to test and calibrate data capture as
well as analysis method. Satisfactory results were generated by analyzing these
stars. Then, the time-series analysis was performed on NSV
13601.\par\vspace{\baselineskip}

\noindent First the  LC(Light Curve) were inspected on different time scales in
Vstar to roughly find  the daily variation of object star. Then the detailed
time-series analysis, using DC-DFT(Date compensated Discreet Fourier Transform)
facility of Vstar was carried out on the MKBU data. It gave the top frequency
hits, which were probable candidates for the object star's period. Periodogram and
Amplitude spectrum were generated and after selecting most probable frequency for
the period of variable star, PLC(Phased Light Curves) were drawn. The PLC and
residual spectrum was  model fit, and mean fit curve were also drawn.  This
sequence was carried out on all MKBU collected data of variable and suspected
variable stars.\par\vspace{\baselineskip}

\noindent In the following sections the respective graphs, figures and tables are
given with each star and in the end their period search results are also given in
respective tables.  Period04 software
package\footnote{\texttt{http://www.univie.ac.at/tops/Period04/}} was also used to
verify the results, (Period04 is very well known for searching frequency of
time-series data using Fourier transform).  Accurate results provided by Period04
were also compared with results of Vstar with their respective photometric filters
and then the frequency which were common in all the results were
selected.\par\vspace{\baselineskip}

\noindent For NSV 13601, scattered data were available from a few sources. The
observations at MKBU have produced new data for these suspected variables. Already
massive efforts are going on to gather and process data from various sources
automatically and update the status.\par\vspace{\baselineskip}

\noindent \textbf{Presentation of Analysed data:}\par\vspace{\baselineskip}

\noindent The following sections present analysis of  time series data from MKBU
Telescope as well as from  other sources (where available) for NSV 13601 using
Vstar. Each presented analysis,   graphically contains several
results from various steps describing either data sufficiency or are checkpoints
in analysis. Descriptions given below are for NSV 13601(Fig 2 to 7). The sequence of analyzed data presented is by VSTAR only. We used PERIOD04's analysis result directly in Table 3 for comparison purpose. This sequence is followed on data from MKBU telescope (for both V and Ic filters). Finally the conclusion drawn are presented.\par\vspace{\baselineskip}

\noindent (Fig. 2) shows light curve(LC) of NSV 13601, it is, Johnson V mag vs JD
graph.
Only in MKBU Data-Ic filter,  LC graph is Cousins Ic mag vs JD. Otherwise for all
other data Johnson V filter mag vs JD graph is used.\par\vspace{\baselineskip}

\noindent On the time-series, DC-DFT was applied and as a result, top-frequency
hits were found along with Periodogram(Fig. 3) and Amplitude spectrum.\par\vspace{\baselineskip}

\noindent Frequency with highest amplitude was selected as the most probable
frequency of  the variable star and phased light curve(PLC)(Fig. 4) were drawn. By
looking at the PLC it can be reasonably inferred whether the star is a variable or
not.\par\vspace{\baselineskip}

\noindent Table - 3 is the results of pre-whitening process, in which every
possible frequency is extracted and then that frequency is removed from the
remaining data and again Fourier transform is applied on residuals, this process
is followed till the residual value reaches near the 0.01. After that it is
assured that no other possible frequency of variable star is left in data
anymore.\par\vspace{\baselineskip}

\noindent Sometimes when time-series data does not perform/pass these checks, as
it should, then Period04's frequency analysis and Vstar's Top frequency hits were compared and the most common frequency was selected for the period.\par\vspace{\baselineskip}

\noindent \textbf{Time-Series Analysis of Suspected Variable Star NSV
13601:}\par\vspace{\baselineskip}

\noindent \textbf{Vstar Analysis:(MKBU Data)}\par\vspace{\baselineskip}

\noindent \textbf{1. Visual Filter (V):}\par\vspace{\baselineskip}

\noindent From Figure 2 to 4, LC, DC-DFT periodogram, PLC are shown respectively for MKBU data in V filter.\par\vspace{\baselineskip}

\noindent \textbf{2. Infrared Filter (Ic):}\par\vspace{\baselineskip}

\noindent From Figure 5 to 7, LC, DC-DFT periodogram, PLC are shown respectively for MKBU data in Ic filter.\par\vspace{\baselineskip}

\clearpage
\newpage

\begin{figure}

\centering{\includegraphics[width=10cm]{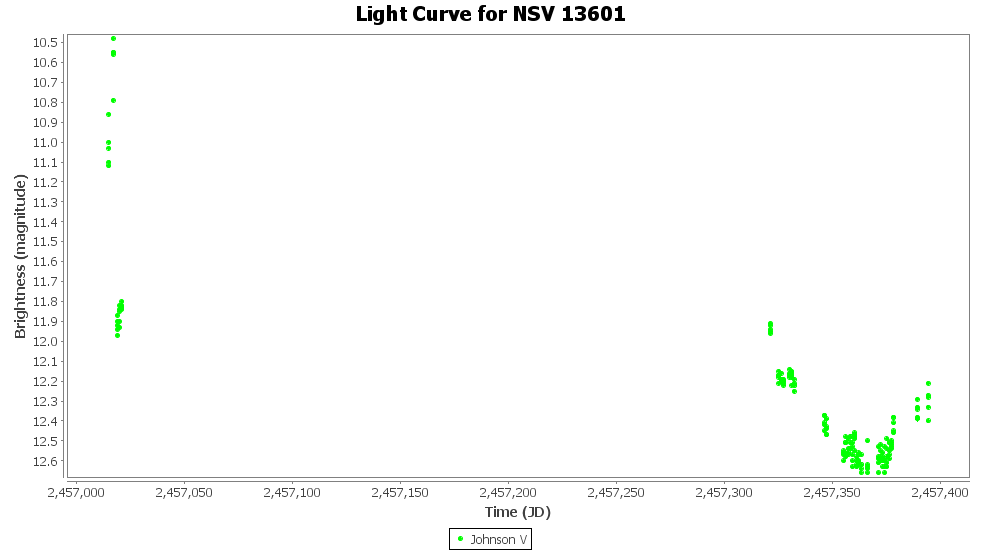}}

\caption{Light curve of NSV 13601 (MKBU Data-V)\label{Fig2}}

\end{figure}

\begin{figure}

\centering{\includegraphics[width=10cm]{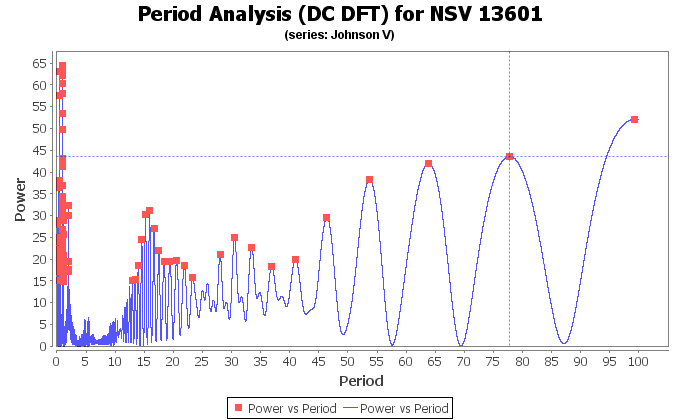}}

\caption{DC-DFT Periodogram of NSV 13601\label{Fig3}}

\end{figure}

\begin{figure}

\centering{\includegraphics[width=10cm]{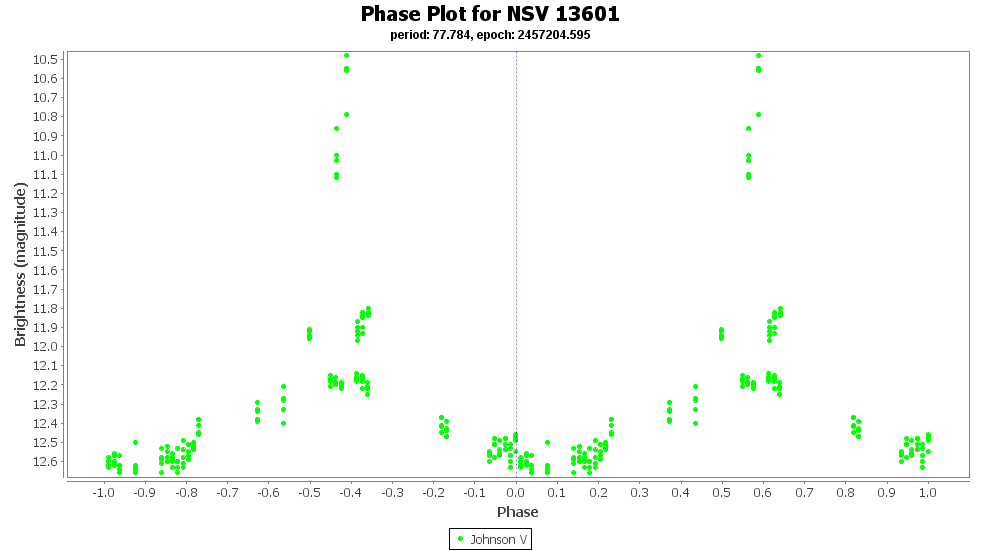}}

\caption{Phased Light Curve of NSV 13601\label{Fig4}}

\end{figure}

\clearpage
\newpage

\begin{figure}

\centering{\includegraphics[width=10cm]{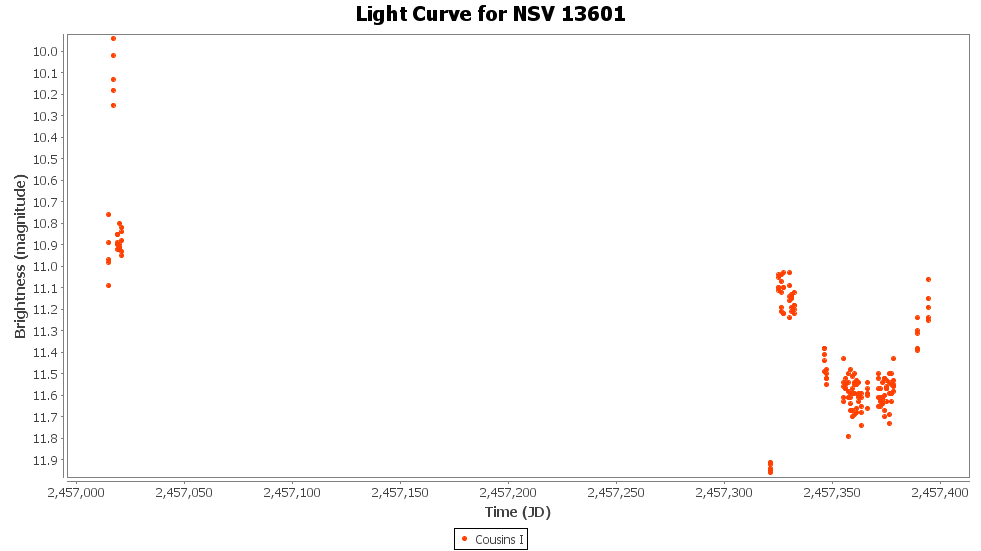}}

\caption{Light curve of NSV 13601 (MKBU Data-Ic)\label{Fig5}}

\end{figure}

\begin{figure}

\centering{\includegraphics[width=10cm]{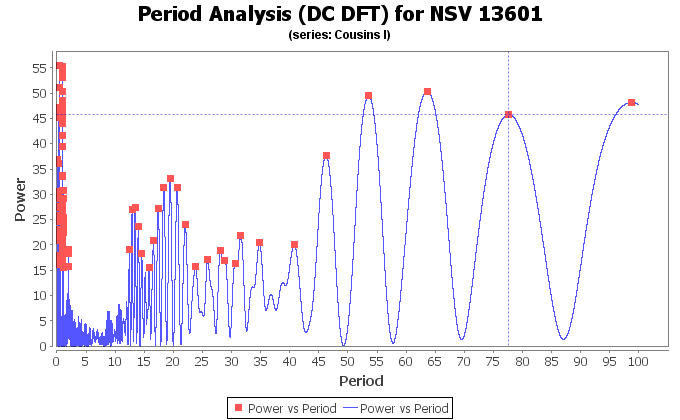}}

\caption{DC-DFT Periodogram of NSV 13601\label{Fig6}}

\end{figure}

\begin{figure}

\centering{\includegraphics[width=10cm]{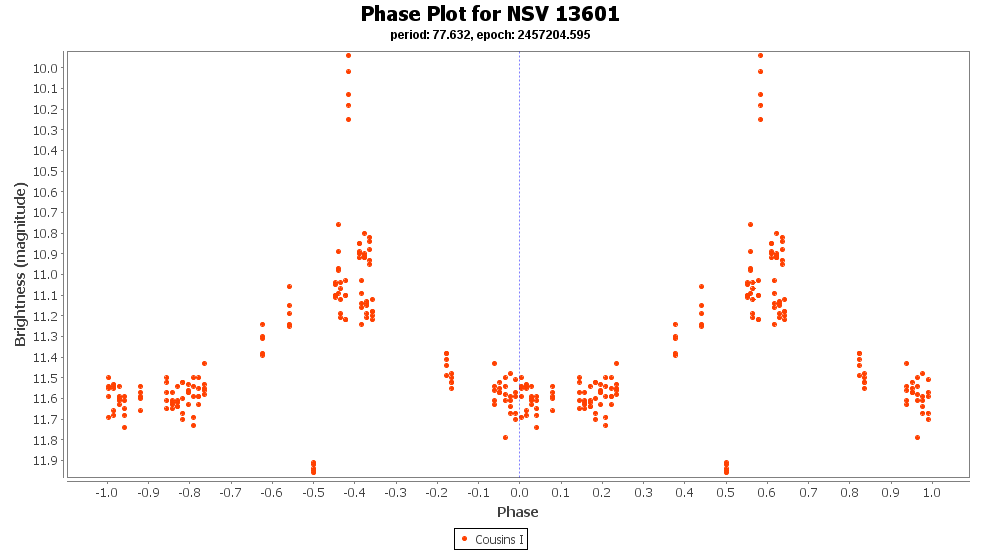}}

\caption{Phased Light Curve of NSV 13601\label{Fig7}}

\end{figure}

\clearpage
\newpage

%
%
%
%
%
%

\clearpage

\section{Results}
\label{sect:res}

Many of the known variable stars were studied for calibration of MKBUs photometric
system and method. Many suspected variable stars were shortlisted and studied by
MKBU telescope. Here presented their results. Their period as well as their
magnitude were determined. Presented below are the summarized results of analysis
of the selected stars.\par\vspace{\baselineskip}

\begin{table}[ht]
\resizebox{\textwidth}{!}{
\begin{minipage}{14cm}
\centering
\begin{tabular}{|l|l|l|}

 \hline
 \textbf{Star Name}&\textbf{MKBU Data(V)}&\textbf{MKBU Data(Ic)}\\

 ($NSV 13601$)&($d$)&($d$)\\

 \hline
 VStar&77.784&77.632\\
 \hline
 Period04&77.058&49.560\\
 \hline
\end{tabular}
\end{minipage}}
\caption{\textbf{Time-series analysis results of NSV 13601}}
\end{table}

\section{Conclusion}
\label{sect:concl}

Conclusion: NSV 13601 is a variable star. Some more study is required to define
it's variability more precisely.\par\vspace{\baselineskip}

\noindent Our aim and strategy was to first standardize our photometric system and
then search for suspected variables, maybe short-period, and do their time-series
analysis. As a result of our study, we found all of our suspected variable star's
periods and magnitude to the scientifically satisfactory
level.\par\vspace{\baselineskip}

\noindent At the time of our selection($\sim$ 2014) for NSV 13601,  it was
classified as a suspected variable status on VSX, CDS and other astronomical
databases.\par\vspace{\baselineskip}

\noindent *Recently(May 15, 2019) it was found that some databases (mainly CDS)
have been updated and are now showing some of earlier suspected stars as variable
stars. VSX is still showing them as suspected as before.(as on 1/06/2019) (sample
screen shot in Fig. 8)\par\vspace{\baselineskip}

\clearpage
\newpage

\begin{figure}

\centering{\includegraphics[width=10cm]{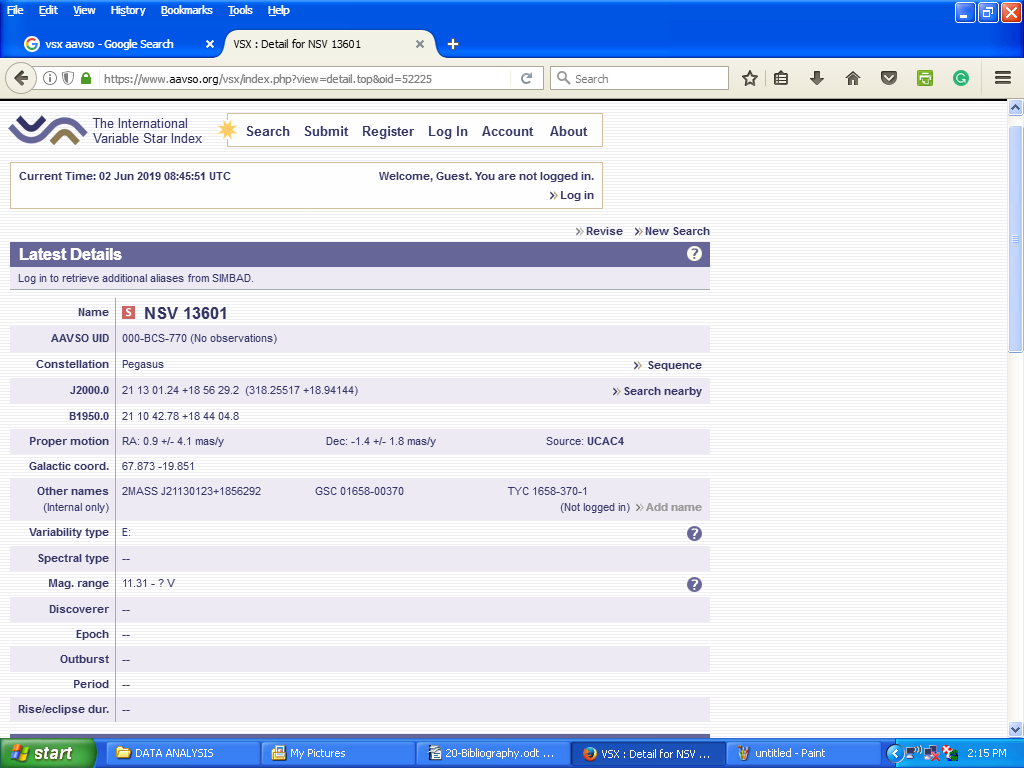}}

\caption{Recent screenshots of NSV 13601 on VSX, still classified as
suspected variable\label{Fig8}}

\end{figure}

\begin{acknowledgements}

We would like to acknowledge the M. K. Bhavnagar University for providing
Astronomical research facility at Kumari Aanya Binoy Gardi Observatory and We
would also like to thank, IUCAA(Inter University Center For Astronomy And
Astrophysics), AAVSO(American Association for Variable Star Observers),
VSX(Variable Star Index)(AAVSO).\par\vspace{\baselineskip}

\end{acknowledgements}

\label{lastpage}


\begin{thebibliography}{99}

    \bibitem[Bhatnagar et al.(2001)]{Bhatnagar+etal+2001} Bhatnagar S.P., Dodia
        U., Anandram M. N., Kagli B. A., Gupta R., 2001, JISI, 31, 234, 239

    \bibitem[Bessell(1990)]{Bessell+1990} Bessell M. S., 1990, PASP, 102, 1181B

    \bibitem[Rodrigez-Gil(2005)]{Rodrigez-Gil+2005} P. Rodrigez-Gil, 2005, A \& A,
        431, 289

    \bibitem[Gohil and Bhatnagar(2019)]{Gohil+Bhatnagar+2019} Gohil V. A.,
        Bhatnagar S. P., 2019, RAA, 19, 9, 125

    \bibitem[Miles(1998J)]{Miles+1998} Miles, R., 1998J, BAA, 108, 65M


\end{thebibliography}
\end{document}